\documentstyle[12pt]{article}

\begin{document}
\def\thebibliography#1{\section*{REFERENCES\markboth
 {REFERENCES}{REFERENCES}}\list
 {[\arabic{enumi}]}{\settowidth\labelwidth{[#1]}\leftmargin\labelwidth
 \advance\leftmargin\labelsep
 \usecounter{enumi}}
 \def\newblock{\hskip .11em plus .33em minus -.07em}
 \sloppy
 \sfcode`\.=1000\relax}
\let\endthebibliography=\endlist

\hoffset = -1truecm
\voffset = -2truecm


\title{\large\bf
Baryon Self-Energy With QQQ Bethe-Salpeter Dynamics In The
Non-Perturbative QCD Regime: n-p Mass Difference
}

\author{
{\normalsize\bf
Anju Sharma
}\\
{\normalsize National Inst. for Adv. Studies, I.I.Sc.Campus,
Bangalore-560012,
India}\and
{\normalsize\bf A.N.Mitra\thanks{e.mail: (1) ganmitra@.nde.vsnl.net.in
(subj:a.n.mitra); (2) anmitra@csec.ernet.in}}\\
\normalsize 244 Tagore Park, Delhi-110009, India
}

\date{6 June 1999}

\maketitle

\begin{abstract}
        A $qqq$ BSE formalism based on $DB{\chi}S$ of an input 4-fermion
Lagrangian of `current' $u,d$ quarks interacting pairwise via
gluon-exchange-
propagator in its {\it non-perturbative} regime, is employed for the
calculation of baryon self-energy via quark-loop integrals. To that end the
baryon-$qqq$ vertex function is derived under Covariant Instantaneity Ansatz
(CIA), using Green's function techniques. This is a 3-body extension of an
earlier $q{\bar q}$ (2-body) result on the exact 3D-4D interconnection for
the respective BS wave functions under 3D kernel support, precalibrated to
both $q{\bar q}$ and $qqq$ spectra plus other observables. The quark loop
integrals for the neutron (n) - proton (p) mass difference receive
contributions from : i) the strong SU(2) effect arising from the $d-u$ mass
difference (4 MeV); ii) the e.m. effect of the respective quark charges. The
resultant $n-p$ difference comes dominantly from $d-u$ effect ($+1.71 Mev$),
which is mildly offset by e.m.effect ($-0.44$), subject to gauge
corrections.
To that end, a general method for QED gauge corrections to an arbitrary
momentum dependent vertex function is outlined, and on on a proportionate
basis from the (two-body) kaon case, the net n-p difference works out at
just
above 1 MeV. A critical comparison is given with QCD sum rules results. 
\end{abstract}

\newpage

\section{Introduction:Relativistic 2- and 3-Quark Hadrons}
\par
        Soon after the advent of the Faddeev theory [1], the
{\it relativistic} 3-body problem [2] attracted instant attention as a
non-trivial {\it dynamical} problem, as distinct from earlier ``kinematical"
attempts [3] at a relativistic formulation of its wave function.
In this respect the relativistic 3-baryon problem had been more of academic
than practical interest (until the `pion' got involved as a key ingredient),
but the  situation changed qualitatively when this 3-body problem started
being viewed at the quark level. Looking back after 25 years it appears that
the first serious attempt in this direction was made by Feynman et al [4]
who  gave  a unified formulation of both the $q{\bar q}$ (meson) and $qqq$
(baryon) problems under a common dynamical framework, bringing out  rather
sharply  an  underlying duality between these two systems  which in turn
signifies a more basic duality between a $qq$ diquark [5] and a ${\bar q}$
antiquark. Indeed the diquark description is quite compact and adequate
for many practical purposes involving the baryon, but the more microscopic
$qqq$ description which brings out the fuller permutation ($S_3$) symmetry
in
the baryon is necessary for the actual details of a full-fledged dynamical
treatment [4].

\subsection{BS Dynamics:3D vs 4D Forms}
\par
        Although the FKR theory [4] marked the first step in this direction,
it suffered from an inadequate treatment of the time-like d.o.f. which
showed up in several ways. The latter has by itself a long history of
attempts
at 3D formulations of the 4D Bethe-Salpeter Equation (BSE) for $q{\bar q}$
$qqq$ systems:Instantaneous Approximation (IA) [6a]; Quasipotentials [6b];
On-shellness of the associated propagators [6c]; and others [6d]. Some of
these approaches have been reviewed elsewhere [7] in the context of a
unified BS formulation of $q{\bar q}$ and $qqq$ in the FKR [4] spirit.
\par
        A different form of 3D reduction, which is of more recent origin
[8,9]
is based on the Markov-Yukawa Transversality Condition [10] in which the BSE
kernel is given a 3D support by demanding that it be a function of the
relative momentum ${\hat q}$ transverse to the total 4-momentum $P$.
This condition, termed the ``Covariant Instaneity Ansatz" (CIA) [8], is
somewhat complementary to the more conventional approaches [6] in which
the BS kernel is left untouched but the propagators are manipulated in
various
ways to give a 3D reduction to the BSE, whereas in the Markov-Yukawa [8-10]
method, which has been termed CIA [8], it is the other way around. In the
conventional 3D reductions [6] there is no going back to the original 4D
form, whereas in the CIA [8], the two forms are fully interchangeable
according to need, thus offering a possible Lorentz covariant way to
reconcile
the apparently conflicting demands of spectroscopy [11] needing a 3D BSE,
with the 4D BS vertex functions needed for quark-loop integrals. The
effectiveness of the CIA in giving a concrete shape to such a ``two-tier"
philosophy of spectra-cum-loop integrals was summarised in a  semi-review
[12]
in the form of appropriate BSE's for  both $q{\bar q}$ and $qqq$ systems
with
vector-type kernels [7] with 3D support, albeit with slight modifications
[7]
in the respective BSE structures to facilitate greater `manoeuvreability',
in
the spirit of similar efforts [13] in the past. Further, the observed
spectroscopy [11] is well satisfied on both $q{\bar q}$ [14] and $qqq$ [15]
sectors with a {\it {common set}} of parameters for the respective kernels
(the $qq$ kernel has just half the strength of the $q{\bar q}$ kernel due to
color effects), so that the respective vertex functions are entirely
determined within the CIA formalism.

\subsection{ QCD-Motivated BSE in 3D-4D Form}
\par
        The other aspect of this `two-tier' formalism concerns the crucial
property of chiral symmetry {\it and} its dynamical breaking. The first part
(chiral symmetry) is ensured without extra charge by the {\it vector}
character of the kernel that had been present all along in this program
[7,16], since the BS-kernel is a direct reflection of an effective 4-fermion
term in the input Lagrangian. Indeed the vector type character of the latter
lends  a natural gluon exchange flavour to such a pairwise interaction among
`current' (almost massless) $u,d$ quarks at the Lagrangian level. This
structure is quite general [17], and can be adapted to the QCD requirements
on the gluonic propagator involved in the pairwise interaction kernel. Of
this, the perturbative part (which is well understood) is quite explicit,
but the {\it non-perturbative} (infrared) part is not yet derivable from
formal QCD premises [18]. It can nevertheless be simulated in a sufficiently
realistic manner at the phenomenological level [7,19], so as to satisfy the
standard constraints of confinement as well as explicit QCD features [20] in
terms of a basically 3D BSE kernel structure.
\par
        The second part, viz., dynamical breaking of chiral symmetry
($DB{\chi}S$) is implemented via the Nambu-Jonalasino mechanism [21] whose
full-fledged form amounts to adopting the `non-trivial' solution of the
Schwinger-Dyson Equation (SDE) derived from a given input, chirally
symmetric
Lagrangian with current quarks. A mass function $m(p)$ [17,18] is thus
generated whose low-momentum value may be identified with the bulk of the
`constituent' mass ($m_q$) of the $u,d$ quarks. This accords with Politzer
additivity [22], viz., $m_q = m(0) + m_c$; where $m_c$, the current mass,
is small. This was also shown in the context of a BSE-cum-SDE treatment [23]
within the CIA formalism [8]. Thus formally the BS-kernel may be regarded as
a {\it non-perturbative} gluon propagator [23] in a BSE framework involving
the dynamical/constituent mass [19, 21-24] in the quark propagator.
\par
        To recapitulate, the CIA which gives an exact interconnection
between
the 3D and 4D forms of the BSE, provides a unified view of 2- and 3-quark
hadrons, its 3D reduction being meant for spectroscopy [14-15], and the
reconstructed 4D form [12,13] for identifying the respective hadron-quark
vertex functions as the key ingredients for 4D quark-loop integrals.
The formalism stems from a strongly QCD-motivated Lagrangian with
{\it current} quarks whose pairwise interaction is mediated by a gluonic
propagator in its non-perturbative regime. The QCD feature of chiral
symmetry
is ensured by the vector nature of this interaction, while its dynamical
breaking is the result of a non-trivial solution of the SDE [17,23]. Thus,
unlike in conventional potential models [25], the constituent mass so
generated
is {\it not} a phenomenological artefact, but the result of a
self-consistent
solution of the SDE [17, 19, 23], so that the standard (constituent) mass
employed for spectroscopy [14-15] `checks' with the output dynamical mass
at low momentum [23]. Thus there are only two genuine input parameters
$C_0$,
$\omega_0$, that characterize the (phenomenological) structure of the
non-perturbative gluon propagator which serves for both the 2- and 3-quark
spectra in a unified fashion [14-15]. In this formalism, these two constants
play a role somewhat similar to that of the (input) `condensates' in the
theory of QCD sum rules [26].

\subsection{Comparison With Chiral Perturbation Theory, Etc}
\par
        Before proceeding further, let us pause to compare this approach
with
other dynamical methods, e.g., chiral perturbation theory [27] which has
more explicit QCD features, albeit in the perturbative regime, leading to
expansions in the momenta. This is a powerful theoretical approach employing
the (chiral) symmetry of QCD; its essential parameters are the current quark
masses, and the method works very efficiently where its premises are
logically applicable. Thus it predicts the {\it {ground state}} spectra of
light quark hadrons, including their mass splittings due to strong and e.m.
breaking of SU(2), but {\it not} the spectra of L-excited hadrons. The
latter
on the other hand demand a ``closed form" approach to incorporate the
``soft"
{\it {off-shell}} effects which in turn require a non-trivial handle on the
{\it infrared} (non-perturbative) part of the gluonic propagator, something
which the present state of the QCD art does not yet provide. Thus one needs
a
phenomenological input even in standard BSE-SDE approaches [18], as
discussed
elsewhere [23]. The chiral perturbation theory [27] also lacks this vital
ingredient, as seen from the absence of form factors in its `point'
Lagrangians [27] with at most derivative terms. This shows up, e.g., through
its inability to predict L-excited spectra, and finer aspects (such as
convergence) of 4D quark-loop integrals which depend crucially on these
``off-shell" features. {\it Physically} this amounts to the absence of a
`confinement scale' which governs these form factors. In other BSE-cum-SDE
approaches [17-19], including the present `two-tier' CIA formalism [8,12],
this `scale' is an integral part of the structure of the non-perturbative
part
of the gluon propagator [19,23], with a built-in QCD feature of chiral
symmetry and its dynamical breaking through the non-trivial solution of the
SDE [17,19,23]. This not only facilitates the prediction of L-excited
spectra
[19,14-15] but also provides a form factor for the hadron-quark vertex
function
which greatly enhances its applicability to various 4D quark-loop integrals;
see [8, 23-24, 28].

\subsection{Application to n-p Mass Difference with 3D-4D $qqq$ BSE}
\par
        After this excursion on the philosophy of this two-tier BSE
approach, vis-a-vis some others [26,27], we may now state the objective of
the present paper: A typical application of the 4D baryon-$qqq$ wave
function
reconstructed [12] from the 3D $qqq$ BSE, to the $n-p$ mass difference, as
a 3-body generalization of the corresponding $q{\bar q}$-meson problem [28].
Unlike the 2-body case, however, where the 3D-4D interconnection is exactly
reversible [8], a 4D reconstruction for a 3-body system involves a loss of
information on the 4D Hilbert space, so that the reversal of steps is in
principle non unique, and requires a 1D $\delta$-function to fill up the
information gap between 3D and 4D Hilbert space which may be directly
attributed to the CIA ansatz of a 3D support to the pairwise kernel. The
2-body case just escapes this pathology as it represents a sort of
degenerate
situation, but the price of a 3D kernel support must show up in a
reconstruction of the 4D BSE from its reduced 3D form in any $(n > 2)$-body
problem [29]. A plausible `CIA' structure for the 4D $qqq$ wave function was
suggested in [12] in a semi-intuitive fashion, but a more formal
mathematical
basis has since been found [29] through the use of Green's function
techniques, so that the reconstructed 4D form reduces exactly to the (known)
3D form as a consistency check [29]. The final result, which is almost the
same as the earlier conjecture, eq.(5.15) of [12], except for a constant
that
does not affect the normalization, contains a 1D $\delta$-function
corresponding to the on-shell propagation of the spectator between two
successive vertex points. As explained in detail in [29], this 1D $\delta$-
function must not be confused with any signature of ``non-connectedness" in
the 3-body wave function [30], since the 3D form is {\it {fully connected}}.
Rather, it can be likened to a  (Fermi-type) $\delta$-function potential in
estimating the effect of chemical binding on the scattering of very slow
neutrons by a hydrogen molecule [31], as a practical devise to fill up the
vast mismatch of scales in the interactions at nuclear vs molecular levels.
In any case the 1D $\delta$-function appearing in this structure is entirely
innocuous as it gets integrated out in any {\it physical} (quark loop)
amplitude including the BS normalization (see Sec.2 below).
\par
        Now to recall the {\it physics} of the n-p mass difference, this
quantity receives contributions of {\it opposite} signs from two main
sources:
i) a positive one from the strong SU(2) $d-u$ mass difference;
ii) a negative one from e.m. splittings. A third source, the effect of quark
condensates, which plays a crucial role in QCD sum rule studies [26] gives
rather small contributions in this non-perturbative approach, as already
found in the meson case [28], and will therefore be neglected; for a
detailed discussion on this issue, see sec.5.2.
\par
        On the other hand the problem of gauge invariance (g.i.) of the e.m.
contribution (ii) is a  more tricky issue in view of the otherwise arbitrary
nature of the extended form factor associated with the $baryon-qqq$ vertex
function, in contrast with the {\it {point-like}} vertices involved, e.g.,
in
the corresponding QCD-sum rule studies [32]. For the same reason, the g.i.
issue could not be addressed in [28] for the meson case, in the hope that it
would be of the same order of magnitude as the relatively small (20
of the e.m. contribution [28]. The g.i. problem with arbitrary hadron-quark
vertex functions is best left to a separate, more substantial investigation.
In the meantime in this paper we shall estimate the g.i. corrections for a
{\it 2-body} problem on the lines of [32] adapted to an arbitrary meson-
quark vertex function. The steps are indicated in an appendix (Appendix C)
for the kaon problem as a test case, the results of which are provisionally
considered as an indication of the nature of the g.i. corrections to be
expected for the 3-body problem on hand, pending a formal treatment later.
The kaon result indicates an increase of 0.612 MeV in the e.m. self-energy
(1.032 MeV) arising from fig 1(b) of [28].

\subsection{Contents of the Paper}
\par
        In Sec.2 we collect the various pieces of the central quantity of
the present investigation, viz., the 4D $baryon-qqq$ vertex function in
terms of 3D quantities, with the inclusion of the spin and isospin d.o.f.
on the lines of an earlier study [33]. Thus equipped, we outline the main
steps leading to an explicit evaluation of the normalization integral, using
Feynman diagrams shown in figs.1(a.b,c). A {\it complex} basis [9, 34, 35]
for 3D momentum variables facilitates the evaluation of the resulting
$3D\times 3D$ integrals, {\it after} the time-like momenta have been
eliminated by `pole' integrations on identical lines to the corresponding
$q{\bar q}$ problem [12,23,24]. In Sec.3 we evaluate the `shift' in the
nucleon mass due to strong SU(2) breaking, by inserting a mass shift
operator
$- \delta m {\tau_3}^{(i)}/2$  in place of $i{\hat \gamma}_\mu e_i$ at each
of
the corresponding $\gamma$- vertices of figs.1(a,b,c), as shown in figs.
2.(a,b,c). Here $\delta m$ = 4 MeV is the `standard' d-u mass difference
[24,28] taken as the basic input.  Sec.4 sketches the evaluation of the
e.m. contribution in accordance with the diagrams of fig.3(a,b,c). The
details of the e.m. approximations employed are collected in Appendix A.
Appendix B sketches the main steps of the derivation [29] for the 4D
structure, eq.(5.15) of [12], of the baryon-$qqq$ vertex function by the
Green's function method for 3 spinless quarks. Finally Appendix C sketckes
the steps leading to the g.i. corrections [32] for the kaon case, as a
sort of facsimile of similar corrections expected for the $n-p$ problem on
hand. Sec.5 summarises our findings and conclusions vis-a-vis other methods.


\section{Normalization of the Baryon-$qqq$ Vertex Function}
\par
        To outline the structure of the baryon-$qqq$ vertex function
from a CIA-governed BSE [12-13], we shall generally follow the notation,
normalization and phase convention for the various symbols as given in [13],
but adapted to the {\it equal} mass kinematics $(m_1 = m_2 = m_3 = m_q)$.
The SU(2) mass difference $\delta m$ $(\approx 4 MeV)$ between $d$ and $u$
quarks will be taken into account only through a 2-point vertex
$[- \delta m {\tau_3}^{(i)} /2]$ inserted in the quark propagators in figs.2
(in place of $i\gamma_\mu e_i$  for a photon), but not in the structure of
the vertex function. The vertex function is written in {\it three} pieces in
each of which {\it one} quark plays the role of the `spectator' by turn. For
the spin structure (not given in [13]) we employ the convention of [3] which
was extended in [33] to incorporate the $S_3$-symmetry for the spin-cum-
isospin structure in the Verde [36] notation [37]. The full 4D BS wave
function $\Psi$ reads as [13,33,34] :
\setcounter{equation}{0}
\renewcommand{\theequation}{2.\arabic{equation}}
\begin{equation}
\Psi \Delta_1 \Delta_2 \Delta_3 = (\Gamma_1 + \Gamma_2 + \Gamma_3) \times
  [\chi' \phi' + \chi" \phi"]/{\sqrt 2};
\end{equation}
\begin{equation}
\Delta_i = {m_q}^2 + {p_i}^2 ; \quad (i = 1,2,3).
\end{equation}
Here $\chi'$ and $\chi"$ are the relativistic ``spin" wave functions
in a 2-component mixed symmetric $S_3$  basis which for a {\bf 56} baryon
go with the associated isospin functions $\phi'$ and $\phi"$ respectively.
These are given by [3,33] :
\begin{equation}
[\chi']_{\beta\gamma;\alpha} = [(M-i\gamma.P) i\gamma_5 C/{\sqrt 2}]_
{\beta\gamma} \times U(P)_\alpha/(2M)
\end{equation}
\begin{equation}
[\chi"]_{\beta\gamma;\alpha} = [(M-i\gamma.P) \gamma_\mu C/{\sqrt 6}]_
{\beta\gamma} \times {i\gamma_5 \gamma_\mu U(P)}_\alpha/(2M)
\end{equation}
in a spinorial basis [3,33] in which the index $\alpha$ refers to the
`active' quark (interacting with an external photon line, fig.1),
while $\beta,\gamma$ characterize the other two, with the further convention
that $\gamma$ refers to the ``spectator" in a given diagram, fig.(1).
The `hat' on  $\gamma$ signifies its perpendicularity to $P_\mu$, viz.,
${\hat \gamma}.P = 0$. The notations in eqs.(2.3-4) are standard, with a
common Dirac basis for the entire structure, and `C' is the charge
conjugation operator for quark $\# 3$ in a 23-grouping [3,33]. $P_\mu$
is the baryon 4-momentum, $U(P)$  is  its  spinor  representation,
and $(M-i\gamma.P)/(2M)$ its energy projection operator [3,33].  Further,
because of the full $S_3$-symmetry of the last factor in (2.1), the
$(1,2,3)$ indices can be permuted as needed for the diagram on hand.
Thus in fig.1a, $\# 1(\alpha)$ interacts with the photon ; $\# 2(\beta)$ is
the quark which has had a `last' $qq$-interaction with $\# 1(\alpha)$
before
emerging from the hadronic `blob', while $\# 3(\gamma)$ is the spectator
[33]. In fig.1b, the roles of $\# 1$ and $\# 2$ are reversed so that, of the
two `active quarks' $\# 1$ and $\# 2$, $\# 2(\alpha)$ now interacts with the
photon, $\# 1(\beta)$ has had the last $qq$- interaction with $\#
2(\alpha)$,
while $\# 3(\gamma)$ still remains the `spectator'. These roles are
cyclically
permuted, with two more such pairs of diagrams, fig.1c), to give an
identical chance to each of the quarks in turn [33]. Thus there are 3 such
pairs of diagrams, of which only one pair is shown. An identical
consideration
applies to figs.2(a,b) with $i\gamma_\mu e_i$  replaced by
$(-\delta m \tau_3 /2)$ consistently. The spatial vertex functions
$\Gamma_i$
are given for i = 3 by [12] :
\begin{equation}
\Gamma_3 = N_B [D_{12} \phi /{2i\pi}]
\times {\sqrt {[2\pi\delta(\Delta_3).\Delta_3]}}
\end{equation}
where $\phi$ is the full, {\it connected} $qqq$ wave function in 3D form,
and
$D_{12}$ is the 3D denominator function of the (12) subsystem . The second
factor represents the effect of the spectator [12] whose {\it inverse}
propagator ${D_F}^{-1}(p_3)$ {\it {off the mass shell}} is just $\Delta_3$,
eq.(2.2). The main steps leading to this unorthodox structure which has been
derived recently via the techniques of Green's functions [29], are sketched
for completeness in Appendix B. As already noted in Sec.1, and again
explained
in Appendix B, its peculiar singularity structure in the form of a
``square-root" of a 1D $\delta$-function stems from the CIA ansatz of a 3D
support to the pairwise interaction kernel, but it is quite harmless as the
former will appear in a {\it linear} form in the transition amplitude
corresponding to any Feynman diagram as in figs.1-2. The complete
expressions
for $D_{12}$ and $\phi$ are given for the equal mass case (with $\# 3$ as
spectator) by (see [12,15]):
\begin{equation}
D_{12} = \Delta_{12} (M-\omega_3); \quad
\Delta_{12} = 2 {\omega_{12}}^2 - M^2 (1-\nu_3)^2/2
\end{equation}
\begin{equation}
{\omega_{12}}^2 = {m_q}^2 + {{\hat q}_{12}}^2; \quad
2{{\hat q}_{12}}^\mu = {{\hat p}_1}^\mu - {{\hat p}_2}^\mu
\end{equation}
\begin{equation}
{\phi} = \ e^{-({{\hat p}_1}^2 + {{\hat p}_2}^2 + {{\hat
p}_3}^2)/{2\beta^2}}
     \equiv \ e^{-\rho/{3\beta^2}}
\end{equation}
(see further below for the definition of $\rho$).
\begin{equation}
{{\hat p}_i}^\mu = {p_i}^\mu + p_i.P P_\mu/M^2; \quad
{{\hat p}_1}^\mu + {{\hat p}_2}^\mu + {{\hat p}_3}^\mu = 0
\end{equation}
\begin{equation}
{\omega_i}^2 = {m_q}^2 + {{\hat p}_i}^2; \quad \nu_3 = \omega_3/M (on shell)
\end{equation}
The $\beta$-parameter is defined sequentially by [14,15]:
\begin{equation}
\beta^4 = {4 \over 9}M {\omega_0}^2 {\bar \alpha}_s(1-m_q/M)^2 (M-<\omega>);
\quad {<\omega>}^2 = {m_q}^2 + 3\beta^2/8
\end{equation}
\begin{equation}
{{\bar \alpha}_s}^{-1} = {\alpha_s}^{-1} - {2M C_0}{(1-m_q/M)^2
\over{M-<\omega>}};
\end{equation}
\begin{equation}
{6\pi \over{\alpha_s}} = 29 {\ln {(M-<\omega>) \over {\Lambda_{QCD}}}};
\end{equation}
\begin{equation}
\Lambda_{QCD} = 200 MeV;\quad \omega_0 = 158 MeV;\quad C_0 = 0.29
\end{equation}
The normalization $N_B$, eq.(2.5), is given in accordance with the Feynman
diagrams 1(a,b) by the 4D integral ( c.f.[33]) :
\begin{eqnarray}
iP_\mu/M & = & \sum_{123} \int d^4q_{12} d^4p_3{{{\Gamma_3}^*\Gamma_3}
\over{2\Delta_2 \Delta_3}}[<\phi'|(23)'(1)'_\mu |\phi'> +
{1\over 3}<\phi"|(23)"_{\nu\lambda}(1)"_{\nu\lambda;\mu}|\phi">] \nonumber
\\
         &   & + (1 \Leftrightarrow 2)
\end{eqnarray}
where the matrix element for fig.1a is organized as a product of two
spin-factors : a `23-element' expressed as a Dirac trace over the indices
$\beta,\gamma$; and a `1-element' (with suppressed index $\alpha$).
The associated isospin functions $\phi$ are shown according to (2.1).
The contribution of fig.1b is shown symbolically by $1 \Leftrightarrow 2$,
while $\sum_{123}$ indicates the sum over all the 3 pairs cyclically. In
representing eq.(2.12) we have dropped `cross-terms' like
${\Gamma_i}^*\Gamma_j$, where $i \neq j$, since the presence of a
$\sqrt \delta$-function in each $\Gamma_i$  ensures that a simultaneous
`on-shell' energy conservation of $i \neq j$ spectators is not possible
[33].
The various pieces of the matrix elements in (2.14) which can be read off
from fig.1a in terms of the spin functions (2.3-4) are as follows:
\begin{equation}
(1)'_{\nu\lambda;\mu} = {\bar U}(P)S_F(p_1)i\gamma_\mu e_1 S_F(p_1) U(P)
\end{equation}
\begin{equation}
i{S_F}^{-1}(p) = m_q + i\gamma.p
\end{equation}
\begin{equation}
(1)"_{\nu\lambda;\mu} = {\bar U}(P) i{\hat \gamma}_\nu \gamma_5 S_F(p_1)
i\gamma_\mu e_1 S_F(p_1) i\gamma_5 {\hat \gamma}_\lambda U(P);
\end{equation}
\begin{equation}
(23)' = Tr[C^{-1}\gamma_5(M-i\gamma.P)(m_q-i\gamma.p_2)(M-i\gamma.P)
\gamma_5(m_q+i\gamma.p_3)C]/{8M^2}
\end{equation}
\begin{equation}
(23)"_{\nu\lambda} = Tr[C^{-1}{\hat
\gamma}_\nu(M-i\gamma.P)(m_q-i\gamma.p_2)
(M-i\gamma.P){\hat \gamma}_\lambda(m_q+i\gamma.p_3)C]/{8M^2}
\end{equation}
The `strength' $e_i$  of the (zero-momentum) `photon' coupling to
the quark line $p_i$  can be chosen in several ways [38]. We take
here the simplest possibility, viz., $e_i = 1/3$ each. The isospin
matrix element is first eliminated according to [39]:
\begin{equation}
<\phi'|1|\phi'> = <\phi"|1|\phi"> = 1
\end{equation}
\begin{equation}
<\phi'|{\tau_3}^{(1)}|\phi'> = -3<\phi"{\tau_3}^{(1)}|\phi"> =
<\tau_3>_{(p,n)}
\end{equation}
Eq.(2.20) suffices for (2.4), while (2.21) will be needed for
the u-d mass difference operator $-\delta m{\tau_3}^{(1)}/2$ ; see Sec.3.
Next, the evaluation of the traces in (2.15-18) is straightforward,
after noting that (2.15-16), {\it after} spin-averaging, are expressible
as traces. The results are
\begin{equation}
(23)' \theta_{\nu\lambda} = (23)"_{\nu\lambda}
 = (m_q + M\nu_2)(m_q + M\nu_3) \theta_{\nu\lambda}
\end{equation}
\begin{equation}
(1)'_\mu \theta_{\nu\lambda} = (1)"_{\nu\lambda;\mu}
 = [2M\nu_1(m_q + M \nu_1) +
\Delta_1]\theta_{\nu\lambda}P_\mu/(M{\Delta_1}^2)
\end{equation}
where $\theta$ is a covariant Kronecker delta w.r.t. $P_\mu$, viz.,
\begin{equation}
\theta_{\nu\lambda} \equiv \theta_{\nu\lambda}
 = \delta_{\nu\lambda} - P_\nu P_\lambda/P^2 ; \quad (P^2 = -M^2)
\end{equation}
Collecting all these results and simplifying we get
\begin{equation}
{N_B}^{-2} = \sum_{123} \int d^3{\hat p}_3 {(m_q +
\omega_3)\over{2\omega_3}}
\times \int d^3{\hat q}_{12} {D_{12}}^2 {\phi}^2 [e_1 I_1 + e_2 I_2]
\end{equation}
\begin{equation}
2i\pi I_1 = \int Md\sigma_{12}[2M\nu_1(m_q + M\nu_1) + \Delta_1]/
(M{\Delta_1}^2 \Delta_2)
\end{equation}
where we have ``cashed" the $\delta(\Delta_3)$-function arising from
$|\Gamma_3|^2$ against the time-like component of $d^4p_3$, and used
the results
\begin{equation}
d^4q_{12} = d^3{\hat q}_{12} M d\sigma_{12}; \quad
\nu_{1,2} = (1 - \nu_3) \pm \sigma_{12}
\end{equation}
The integration over $d\sigma_{12}$ involves single and double poles arising
from the propagators ${\Delta_{1,2}}^{-1}$ in (2.26), while the value of
$\nu_3$ is taken `on-shell' at $\omega_3 /M$ after the $\delta(\Delta_3)$-
function has been cashed. The result of a basic $\sigma_{12}$-integration is
\begin{equation}
\int M d\sigma_{12} {\Delta_1}^{-1} {\Delta_2}^{-1} = 2i\pi/D_{12}
\end{equation}
from which others can be deduced by differentiation under {\it unequal}
mass kinematics, or directly through a `double pole' integration. The net
result for $I_1 + I_2$, eq.(2.26), is given in eq.(2.41) below. Further,
the individual terms of the summation $\sum_{123}$ in (2.25) are fixed by
the values chosen for $e_i$ (which need not be specified in advance, as they
can be adapted to other conventions too [38]; see Sec.3).
\par
        The integration in (2.25) can be considerably simplified in  a
complex basis [15,34] defined (in momentum space) by :
\begin{equation}
{\sqrt 2} z_i = \xi_i + i\eta_i; \quad {\sqrt 2} {z_i}^* = \xi_i - i\eta_i;
\end{equation}
\begin{equation}
{\sqrt 3}\xi_i = p_{1i} - p_{2i}; \quad 3\eta_i = -2p_{3i} + p_{1i} +
p_{2i};
\end{equation}
where we now employ the alternative notation $p_{1i}$ for ${\hat p}_1^\mu$,
in view of its basically 3D content. In terms of $z_i$  and ${z_i}^*$, the
6D integration in (2.25) is expressed as
\begin{equation}
d^3{\hat p}_3 d^3{\hat q}_{12} = ({\sqrt 3}/2)^3 d^3 \xi d^3 \eta = d^3z
d^3z^*
\end{equation}
The further representation [15,34]
\begin{equation}
d^3z d^3z^* = ({dz_+}{dz_-}^*) ({dz_-}{dz_+}^*)({dz_3}{dz_3}^*)
\end{equation}
where
\begin{equation}
{\sqrt 2}z_+ = R_1 \ e^{i\theta_1};
\quad {\sqrt 2}{z_-}^* = R_1 \ e^{-i\theta_1}
\end{equation}
\begin{equation}
{\sqrt 2}z_- = R_2 \ e^{i\theta_2};
\quad {\sqrt 2}{z_+}^* = R_2 \ e^{-i\theta_2}
\end{equation}
\begin{equation}
{\sqrt 2}z_3 = R_3 \ e^{i\theta_3};
\quad {\sqrt 2}{z_3}^* = R_3 \ e^{-i\theta_3}
\end{equation}
reduces the 6D integration (2.32)  merely to
$\pi^3 d{R_1}^2 d{R_2}^2 d{R_3}^2$, since the $\theta_i$ -variables
({\it not} Euler angles!) are not involved in the integrands encountered,
and
just sum up to $(2\pi)^3$.  The positive variables $R_i$, (i = 1,2,3), are
related to the $\xi_i,\eta_i$ variables by
\begin{equation}
\rho \equiv R_1^2 + R_2^2 + R_3^2 = \xi^2 + \eta^2 = 2 z_i {z_i}^*
\end{equation}
\par
        To convert the variables $\omega_i$ that appear in the integrals
(2.28) in terms of the $R_{1,2,3}$  variables is a straightforward but
tedious process which can be somewhat simplified in terms of the
intermediate variables $\xi^2 - \eta^2 $ and $2 \xi.\eta$ which form a [2,1]
representation [36] of $S_3$-symmetry at the `quadratic' level. Now because
of the full $S_3$ -symmetry of the 6D integral (2.32), together with the
(fortunate) circumstance of equal mass quarks in the problem on hand, the
integrand as a whole is $S_3$ -symmetric which permits the following
simplification: Each of the quantities ${\hat p}_i^2$  and ${\hat q}_i^2$
inside (2.32) can be expanded as
\begin{equation}
{{\hat p}_{1,2}}^2 = \rho/2 + (\xi^2 - \eta^2)/4 \pm {\sqrt 3}\xi/\eta/2;
\quad {{\hat p}_3}^2 = \rho/2 - (\xi^2 - \eta^2)/2
\end{equation}
\begin{equation}
{{\hat q}_{12}}^2 = 3\xi^2/4 = \rho/2 + (\xi^2 - \eta^2)/4
\end{equation}
In all these terms the principal quantity is $\rho/2$, while the
resultant effects of the mixed-symmetric corrections will show up
only in the {\it fourth} order, etc. In the present case of equal mass
kinematics it is a good approximation to neglect the latter terms, as has
also been found for the qqq mass spectral results [15], so that all
quantities are expressed in terms of $\rho$ only :
\begin{equation}
\omega_{1,2,3} \approx \omega_{12} \approx \omega_\rho;
\quad {\omega_\rho}^2 \equiv m_q^2 + \rho/2
\end{equation}
\begin{equation}
D_{12} \approx 2(M-\omega_\rho)[\omega_\rho^2 - (M-\omega_\rho)^2/4].
\end{equation}
The rest of the integration is expressed entirely in terms of the
$\rho$-variable, with the resultant 6D measure given by
\begin{equation}
\int d^3{\hat p}_3 d^3{\hat q}_{12} F(\rho)
= (\pi{\sqrt 3}/2)^3 \int \rho^2 d\rho/2 F(\rho)
\end{equation}
These considerations suffice for evaluating the integrals $I_1$
and $I_2$   whose resultant value is now given for $e_i = 1/3$ by :
\begin{equation}
D_{12}^2 (I_1 + I_2) = [m_q^2 + m_q(M-\omega_\rho) + (M-\omega_\rho)^2/4]
\times (M-\omega_\rho)^3/\omega_\rho + D_{12} (2m_q + M - \omega_\rho)
\end{equation}
Substitution in (2.25) yields $N_B$ directly. The numerical values
are given collectvely at the end of Sec.4.
\section{Strong SU(2) Mass Difference for the Nucleon}
\par
        This calculation is on almost identical lines to Sec.2, except
for the substitution $ie_1 \gamma_\mu$ to $-\delta m {\tau_3}^{(1)}/2$ in
figs.1(a,b) to give figs.2(a,b) which represent the effect of insertion of
a 2-point vertex in a quark line. Indeed we can directly start from the
counterpart of eq.(2.15) which gives the `strong' mass shift as:
\setcounter{equation}{0}
\renewcommand{\theequation}{3.\arabic{equation}}
\begin{eqnarray}
i\delta M_{st} & = & \sum_{123} \int d^4q_{12}d^4p_3 {{\Gamma_3}^* \Gamma_3
\over {2\Delta_2\Delta_3}}\times [<\phi'|(23)'(1)'|\phi'> +
{1\over 3}<\phi"|{(23)"}_{\nu\lambda}{(1)"}_{\nu\lambda}|\phi">]  \nonumber
\\
               &   & + (1 \Leftrightarrow 2)
\end{eqnarray}
where we have now employed eq.(2.21) for the isospin factors, and the
counterparts of (2.16) and (2.18) are respectively
\begin{equation}
(1)' = {\bar U}(P)S_F(p_1)[-\delta m {{\tau_3}^{(1)}}/2]S_F(p_1) U(P);
\end{equation}
\begin{equation}
(1)"_{\nu\lambda} = {\bar U}(P)i{\hat \gamma}_\nu \gamma_5 S_F(p_1)
[-\delta m {{\tau_3}^{(1)}}/2]S_F(p_1) i\gamma_5 {\hat \gamma}_\lambda U(P)
\end{equation}
while the definitions (2.19) and (2.20) remain unaltered. As a result,
eq.(2.24) remains valid, while the counterpart of (2.23) becomes
\begin{equation}
(1)'\theta_{\nu\lambda} = -3 (1)"_{\nu\lambda}
= [2m_q(m_q + M \nu_1) - \Delta_1]\theta_{\nu\lambda}(-\delta
m/2)/\Delta_1^2
\end{equation}
Carrying out the $d\sigma_{12}$-integration, the result for $\delta M_{st}$
is now given by the counterpart of (2.26), viz.,
\begin{equation}
\delta M_{st} = 3N_B^2 \int d^3{\hat p}_3{(m_q + \omega_3)\over {2\omega_3}}
\times \int d^3{\hat q}_{12} {D_{12}}^2 \phi^2 [J_1 + J_2](-\delta
m\tau_3/6)
\end{equation}
in the form of an isospin operator ``$\tau_3$" for the {\it nucleon}, where
we have represented the effect of $\sum_{123}$ by a factor of ``3", and
\begin{eqnarray}
D_{12}^2 [J_1 + J_2] & = & {m_q\over {\omega_\rho}} [(m_q + {1\over 2}
(M-\omega_\rho))(M-\omega_\rho)[2\omega_\rho^2 + m_q(M-\omega_\rho)]
\nonumber \\
                     &   & +(M-\omega_\rho)^2[m_q + (M-\omega_\rho)/2]^2
+ \Delta_{12}(m_q^2-\omega_\rho^2) \nonumber \\
                     &   & + (m-\omega_\rho)(m_q +
(M-\omega_\rho)/2)(\omega_\rho^2 + m_q(M-\omega_\rho)/2) \nonumber \\
                     &   & + \Delta_{12}^2
(1 -m_q(M-\omega_\rho)/\omega_\rho^2)/2]
\end{eqnarray}
as the exact counterpart of (2.43) under the same approximation. It is seen
from (3.5) that the {\it difference} $(n-p)$ is {\it positive}.
\section{E.M. Mass Difference for the Nucleon}
\par
        The diagrams for the e.m. mass difference are given by figs.3
(I,II,III) for a proton ($uud$) configuration to illustrate the
underlying topology in accordance with the roles of the `active' and
`spectator' quarks in turn, as explained in Sec.2. In each of these
diagrams,
two internal quark lines are joined by a photon line. The e.m. vertex at
quark $\# i$ has the strength $e[1 + 3\tau_3^{(i)}]/6$ from  which the
isospin
matrix elements of a product of {\it two} such factors (shown for fig.3.III)
have the forms
\setcounter{equation}{0}
\renewcommand{\theequation}{4.\arabic{equation}}
\begin{equation}
<\phi';\phi"|{(1+3\tau_3^{(1)})/6} \times {(1+3\tau_3^{(1)})/6}|\phi';\phi">
\end{equation}
for the proton $(uud)$ cnfiguration shown in III with $\# 3$  as spectator,
but
in a basis (1;23) (which is consistent with the spin basis, eqs.(2.3-4)),
corresponding to fig.1a, viz.[37,39]:
\begin{equation}
|\phi'> = u_1(u_2d_3-u_3d_2)/{\sqrt 2}; \quad
|\phi"> = (-2d_1u_2u_3 + u_1d_2u_3 + u_1u_2d_3)/{\sqrt 6}
\end{equation}
We note in parentheses that in fig.3.III, the interchange of the two
`active'
quarks $\# 1$ and $\# 2$ does {\it not} give a new configuration, {\it
unlike} in
figs.1 and 2; ((a) versus (b) configurations).
\par
        It is now easy to check that the matrix elements $< >'$  and $< >"$
of (4.1) are $1/9$ and $-1/9$ for the {\it proton} configuration. After
doing
the corresponding neutron case, the two results may be combined in the
single
operator forms [39]:
\begin{equation}
<.>' = (1 + 3\tau_3)/36; \quad <.>" = (1 - 5\tau_3)/36
\end{equation}
where $\tau_3$ is the isospin operator for the nucleon as a whole [see
eq.(3.5)], to be sandwiched between the neutron and proton states. The
resultant isospin factor is then
\begin{equation}
e^2[<.>' + <.>"]/2 = e^2(1-\tau_3)/36 \Rightarrow -e^2\tau_3/36
\end{equation}
After this book-keeping on the charge factors we can drop the isospin d.o.f.
$|\phi>$ from the $qqq$ wave function and, on the basis of the equality of
the $(.)'$ and $(.)"$ contributions (2.22-23) for the spin matrix elements,
it is enough to work with the $(.)'$ type to represent the full effect.
Collecting these details, the net isospin contribution to the e.m. $(n-p)$
mass difference is just $e^2/18$, which (of course) comes out with the
{\it correct} (negative) sign in the resultant e.m.contribution to the total
$n-p$ difference after all the phase factors in the orbital-cum-spin space
have been taken into account. The complete e.m. self energy of the nucleon
(with operator $\tau_3$), with  fig.3.III as the prototype, is now given by
\begin{equation}
\delta M^\gamma = \sum_{123} [-e^2 \tau_3/36]/(2\pi)^4 \int d^4p_3d^4q_{12}
d^4{q_{12}}'\Gamma_3^* \Gamma_3' k^{-2} \times [23]_\mu'[1]_\mu'/
(\Delta_3 \Delta_2 \Delta_2')
\end{equation}
where the various momentum symbols are as shown in fig.3, with the primed
quantities referring to the vertex on the  right, but otherwise written in
the same convention as in eqs.(2.5-10). The symbols within square brackets
are analogous to (2.16-19):
\begin{equation}
[1]_\mu' = {\bar U}(P')S_F(p_1')i\gamma_\mu S_F(p_1)U(P); \quad (P'=P)
\end{equation}
\begin{equation}
[23]_\mu' = {Tr\over
{8M^2}}[C^{-1}\gamma_5(M-i\gamma.P')(m_q-i\gamma.p_2')i\gamma_\mu
(m_q-i\gamma.p_2)(M-i\gamma.P)\gamma_5 (m_q+i\gamma.p_3) C]
\end{equation}
And the product of (4.6) and (4.7) works out as
\begin{eqnarray}
ME & \equiv & {(m_q + \omega_3)\over {\Delta_1 \Delta_1'}}[-(\Delta_1 +
\Delta_1' - k^2)(\Delta_2 + \Delta_2' -k^2)/4 \nonumber \\
   &        & -(\Delta_1 + \Delta_1'-k^2)(m_q\omega_2 + m_q\omega_2'+
2\omega_2\omega_2')/2 \nonumber \\
   &        & - (\Delta_2 + \Delta_2'-k^2)(m_q\omega_1 + m_q\omega_1'+
2\omega_1\omega_1')/2 \nonumber \\
   &        & -(\Delta_1 + \Delta_2)(m_q + \omega_1')(m_q + \omega_2')/2
- (\Delta_1' + \Delta_2')(m_q + \omega_1)(m_q + \omega_2)/2 \nonumber \\
   &        & -(\Delta_1' + \Delta_2)(m_q + \omega_1)(m_q + \omega_2')/2
-(\Delta_1 + \Delta_2')(m_q + \omega_1')(m_q + \omega_2)/2 \nonumber \\
   &        & +(m_q^2 +{1\over 2}(P-p_3-k)^2)[(m_q + \omega_1)(m_q +
\omega_2')
+ (m_q + \omega_1')(m_q + \omega_2) \nonumber \\
   &        & + (m_q + \omega_1)(m_q + \omega_2) + (m_q + \omega_1')(m_q +
\omega_2')]]
\end{eqnarray}
\par
        Some features of this ``master" expression may be noted. There is
a `natural factorization'in the variables $q_{12}$  and $q_{12}'$, except
for the photon propagator $k^{-2}$, $(k = q_{12} - q_{12}')$.  Further,
the two blobs are connected together by the `spectator'variable $p_3$ which
is on the mass shell due to the presence of $\Gamma_3^*\Gamma_3'$ in
eq.(4.3).
\par
        The time-like (pole) integrations over each of $d\sigma_{12}$ and
$\sigma_{12}'$ can be carried out {\it exactly} a la (2.28) and its
derivatives, since the 3D vertex function $D_{12}\phi$  in $\Gamma_3$ does
not involve $\sigma_{12}$, etc. After this step  ${\hat q}_{12}$,
${\hat q}_{12}'$ and ${\hat p}_3$  are the `right' 3D variables for the
`triple integration' whose essential logic may be stated as follows. The
main strategy is to decouple the ${\hat q}$ and ${\hat q}'$ variables
from the photon propagator k  through the following device [28]:
\par
        Since $k$ is basically space- like, it is a good approximation to
replace $k^{-2}$ by ${\hat k}^{-2}$  which equals $({\hat q}_{12} -
{\hat q}_{12}')^2$, and drop  the angular correlation in the two ${\hat q}$-
momenta (since the error in this neglect is zero in the first order [28]).
Next we use the inequality [28]
\begin{equation}
(a^2 +b^2)^{-1} \leq (2ab)^{-1}; \quad a  \rightarrow |{\hat q}_{12}|, etc
\end{equation}
which ensures the necessary factorizability in the q-variables.In principle
the corrections to this inequality can be calculated since the neglected
term
is  approximately equal to $-(a-b)^2/(4a^2b^2)$  which is still
factorizable,
but this refinement is unnecessary in view of the smallness of the e.m.
effect
itself. After this simplification the rest of the integration procedure is
straightforward  since the ${\hat q}$ and ${\hat q}'$ integrations can be
done analytically, and only a 1D integration over $|p_3|$ remains for
numerical evaluation. The necessary expressions are collected in  Appendix A
and the numerical results for all contributions are given as under.
\par
        The key parameters are the quark mass $m_q$  and the size parameter
$\beta^2$, the latter being determined dynamically through the chain of
eqs.(2.11-14). As noted in Sec.1 already, the mass $m_q$  which is usually
called the `constituent' mass, should be viewed as the sum of the (flavour
independent) `mass function' m(p) for small p, plus a small ``current mass"
$m_c$, in the spirit of Politzer additivity [22]. The mass function m(p) was
generated  in this BSE-cum-SDE framework through a  Dynamical Chiral
Symmetry
Breaking mechanism in a non perturbative fashion [23]. Also from some
related
quark-loop calculations with $q{\bar q}$ mesons in recent times [24,28], it
was found that for such `low energy' processes the mass function m(p) is
rather well approximated by m(0), so that [22], $m_q = m(0)+m_c$. Therefore
the $d-u$ mass difference is the {\it same} at the `constituent' or at the
`current' levels, and this is what has been denoted by $\delta m$ in the
text
(figs.2). Its smallness compared to $m_q$  justifies its neglect in all the
functions except where it appears explicitly, viz., fig.2. We take its value
at $\delta m = 4 MeV$, as in related calculations [24,28], while the other
quantities are predetermined from $q{\bar q}$ [14] and $qqq$ [15]
spectroscopy:
\begin{equation}
m_q = 265 MeV; \quad \beta^2 (N) = 0.052 GeV^2
\end{equation}
so that there are {\it {no free}} parameters in the entire calculation.
The results from Secs.2-4 are now summarized for (n-p) as :
\begin{equation}
N_B^{-2} = 5.5209 \times 10^{-4} GeV^{-10}; \quad [e_i = 1/3]
\end{equation}
\begin{equation}
\delta M_{st} = +1.7134 MeV ; \quad \delta M^{\gamma} = -0.4396 MeV.
\end{equation}
Hence
\begin{equation}
\delta M(net) = + 1.28 MeV; \quad (vs. 1.29 MeV : Expt)
\end{equation}
which is the principal result of this investigation, but subject to possible
gauge invariance corrections [32]; see Appendix C and discussion in sec.5.3.

\section {Discussion, Summary and Conclusion}
\par
        This calculation fills up an important gap in the two-tier BSE
formalism under 3D kernel support (CIA) for a simultaneous investigation
of spectra and transition amplitudes of both $q{\bar q}$ and $qqq$ varieties
under a single umbrella [8,12].

\subsection{Recapitulation}
\par
        To recapitulate the main points, the (first stage) 3D reductions of
both the 2-body and 3-body BSE's had yielded good agreement with the
respective spectra [14,15], with a common set of parameters $C_0 = 0.27$
and $\omega_0 = 158 MeV$ characterizing the non-perturbative gluon
propagator,
since the constituent mass $m_q$ for spectroscopy [14,15], is essentially
the dynamical mass function $m(p)$ in the low momentum limit [22, 23].
\par
        More substantial tests of the formalism have come from the (second
stage) reconstruction of the 4D hadron-quark vertex function which carries
the
non-perturbative off-shell information in a {\it closed} form. This exercise
was initially confined to the meson-$q{\bar q}$ vertex function whose exact
reconstruction [8] had led to several useful results from 4D loop integrals
for hadronic and e.m. transition amplitudes [8,16], to like integrals
probing
the momentum dependence of the quark mass function $m(p)$ which is the
`chiral' limit ($M_\pi = 0$) [17,21,23] of the pion-quark vertex function.
Indeed $m(p)$ acts as the form factor for loop integrals determining the
vacuum to vacuum transitions, and is found to predict correctly several
condensates, from the basic $<q{\bar q}>$ [23] to `induced' condensates
[40],
{\it {under one roof}}. Further tests of the hadron-quark vertex function
have
come from SU(2) breaking effects like $\rho-\omega$ mixing [24] and mass
splittings in pseudoscalar mesons [28], with only one additional parameter
representing the $d-u$ mass difference.
\par
        The last link in our two-tier formalism has been a reconstruction
of the 4D baryon-$qqq$ vertex function on $q{\bar q}$ lines [8], to evaluate
3-body loop integrals. Although conjectured some time ago [12], a rigorous
derivation [29] via Green's functions is outlined in Appendix B, in
preparation for our main task: the $n-p$ mass difference on identical
lines to the $q{\bar q}$ case [28], viz., strong SU(2) breaking (fig 2) and
e.m.contribution (fig 3), since the $qqq$ vertex function is entirely
determined by the same (gluon exchange) dynamics [7,23] as $q{\bar q}$.
Strong SU(2) breaking is due to $d-u$ mass difference, a la Weinberg [41],
and more references on the physics of the problem may be found in [28]. The
point to stress is that no free parameters are involved, so that the
final value (4.13), although a single number, must {\it not} be treated as
an isolated quantity, but as an integral part of a much bigger package.
\par
        The `QCD' status of this 3D-4D BSE formalism [8] viv-a-vis chiral
perturbation theory [27] has already been explained in Sec.1: the gluon
exchange character of the pairwise $q{\bar q}$ or $qq$ interactions lends
them a natural chiral invariance property at the input Lagrangian level with
`current' quarks. Thus the `constituent' mass is {\it not} an input, but
emerges as the low momentum limit of the dynamical mass function $m(p)$ that
characterizes the quark propagators appearing in the 2- and 3- body BSE's,
as a result of $DB{\chi}S$ [21, 17, 19, 23], since the `current' masses of
$u,d$ quarks give only a small additive contribution [22]. The empirical
aspect of the gluon propagator concerns only its {\it {non-perturbative}}
regime which often requires separate parametrization even in more orthodox
BSE-SDE formulations [18]. In the present formulation, its explicit
parametrization with two constants $C_0$ and $\omega_0$) [23] symbolizes
a `closed form' representation of non-perturbative effects in the derived
hadron-quark vertex function, but the returns are rich, especially the
interlinkage of 4D loop integrals for different transition amplitudes
[12,23,24,28], with the 3D BSE structures relevant to spectroscopy [14,15].
\par
        In contrast, Chiral Perturbation Theory [27] has an explicit QCD
content, but relies more heavily on a perturbative treatment, as revealed
by expansions in powers of small momenta and ``current" masses $m_c$ [27]
for a systemic derivation of the low energy structure of the Green's
function
in QCD [27]. It is a powerful method, highly successful in predicting items
like ground state masses {\it {as well as}} their splittings, but its lack
of
a closed form representation prevents an equally successful prediction of
`soft' QCD effects in enough details, such as the momentum dependence of the
mass function, or of hadron-quark vertex functions in general, and other
observable effects such as {\it L-excited} spectra.

\subsection{Comparison with QCD-Sum Rules}
\par
        For a comparison with QCD-Sum Rules, it is useful to start by
mentioning our neglect of the condensate contribution inserted in the
internal quark lines, vide Fig 1c  of [28], on the ground that it was found
to be small in the 2-body case within the same BSE framework [28]. This
contrasts sharply with the corresponding QCD-SR scenario [26] wherein
the condensate contributions are the primary source of non-perturbative
effects,  as confirmed by explicit calculations [42]. Even more surprising
is the {\it inversion} of the effect of the $d-u$ mass difference on the
$n-p$ mass difference vis-a-vis the traditional low energy wisdom which
requires $d-u>0$ to make $n-p>0$, as was the original motivation behind
the famous Weinberg proposal [41]. (Incidentally our BSE result conforms to
the Weinberg picture [41] and hence opposite of QCD-SR [42]). Indeed QCD-SR
must rely heavily on the condensate contributions to compensate for the
(negative) effect of the $d-u$ mass difference, and leave a balance [42].
The e.m. effect too makes a comparable contribution to QCD-SR [42]. On the
other hand the BSE approach, eq.(4.12), seems to predict only a modest
e.m. effect which is about a fourth of, and of opposite sign to, the strong
SU(2) breaking effect which dominates the entire scenario.
\par
        What could account for such a sharp division between the two
approaches? Perhaps the reason should be sought in the very difference
in the philosophy behind their respective premises: QCD-SR is basically a
perturbative QCD approach designed from the high energy end, with the
`twist' terms (condensates) representing non-perturbative corrections of
successively higher orders. In contrast the BSE is an intrinsically
non-perturbative approach built from the low energy (spectroscopy) end,
with the hard gluon exchange added perturbatively. The role of condensates
in BSE is effectively subsumed in the constituent (non-perturbative) quark
mass as well as the hadron-quark vertex function, so that any further
condensate effects (fig 1c of [28]) in such a non-perturbative scenario can
at best be residual [28]. This is not the case in QCD-SR wherein the
quark condensates with their isospin splittings are the dominant
source of non-perturbative effects on the $n-p$ splitting [42]. The two
scenarios are thus largely complementary, QCD-SR being rooted in hard
QCD, and BSE-SDE in soft QCD premises respectively.

\subsection{Problem of Gauge Invariance of E.M. Effects}
\par
        Finally we come to a more vulnerable aspect of this investigation
in company with the earlier study [28], viz., the lack of gauge invariance
of the e.m. contribution (fig.(3)). Mercifully the e.m. contributions in
both the meson [28] and baryon (present) cases are about a fourth of the
$u-d$ effect so as hopefully not to upset the overall stability of our
result (4.13), but the need for a proper assessment of the g.i. corrections
can hardly be overestimated. A general method for g.i. two-point functions
for $Q{\bar q}$ systems has been given in [32], but it is not directly
adaptable to extended vertex functions with arbitrary form factors, such as
in the present situation. To handle these structures requires a different
kind of strategy, a full-fledged formulation of which, is beyond the scope
of this (already long) paper, and is best left to a separate communication.
Nevertheless, as noted in Sec.1, we have made a beginning in this paper by
outlining the main steps of the derivation of g.i. corrections for a typical
two-body (kaon) case, in a simple and straightforward fashion, which amounts
to the replacement of various momenta $p_i$ involved in the hadron-quark
vertex functions by $(p_i-e_iA)$, and expanding in powers of the e.m. field,
with a view to calculate the additional diagrams a la [32]. Hopefully this
method of generating e.m. gauge corrections is general enough to apply to
other situations in which arbitrary momentum dependent hadron-quark vertex
functions are involved. For the present situation of QED gauge corrections,
the main steps are sketched in Appendix C, and the resultant correction to
fig 1b of [28] (i.e. fig 1a of [32])  is estimated to be
$-0.612 MeV$, which is $3/5$ times, and of the same sign as, the e.m. value
$-1.032 MeV$ for the mass difference $ K^- - {\bar K^0}$ [28]. If this
result
is taken as a rough indication of the g.i. effect expected for the nucleon
case on hand, it would mean a downward revision of the value (4.13) to about
$1MeV$. However this is only a provisional estimate, pending a regular $qqq$
calculation in the future.

\subsection{Conclusion}
\par
        To conclude, the principal motivation for this investigation, has
been to demonstrate the practical feasibility of such realistic quark-loop
calculations for the relativistic 3-quark problem within a full-fledged (BS)
dynamical framework whose basic parameters are linked all the way to
spectroscopy. The present calculation indeed suggests that not
only quark-loops involving mesons [8,23,24,28] but even those involving the
(less trivial) $qqq$ baryon are amenable to a similar degree of dynamical
sophistication without excessive efforts, so that it makes sense to speak of
an effective ``4-fermion coupling" for both $q{\bar q}$ and $qq$ pairs
within
a common parametric framework. This is somewhat remiscient of Bethe's
``second principle" theory, originally suggested at the two-nucleon level of
nuclear forces, now reinterpreted at the quark level, with a simple
extension
to include the antiquark in the dynamical description. (This extension would
not make sense at the $NN$ level, since the $NN$ and $N{\bar N}$ forces are
very different from each other). Indeed such a dynamics had been strongly
suggested (with concrete examples) in a perspective review not too long ago
[43], but it seemed to have gone largely by default, as evidenced by a
strong
tendency in the contemporary literature to continue to rely on ``ad-hoc form
factors" [44] to simulate the vertex functions, instead of generating them
dynamically. Hopefully, some efforts in this direction have been recently in
evidence [45], using the Nambu Jonalasino model [21] of contact 4-fermion
interactions (although such contact interaction models are probably too
simple to realistically simulate confinement [29]). It is to be hoped that
Bethe's ``second principle" perspective will be upheld by such
investigations,
until such time as a fully satisfactory solution to QCD is  forthcoming.
\par
        The initial draft of this paper was prepared at the National
Institute of Advanced Studies. We are grateful to Dr.Raja Ramanna for the
warm NIAS hospitality. We also acknowledge Ms Chandana's help with some
`difficult' figures. One of us (ANM) is grateful to Prof. S.R.Chaudhury
for a critical discussion on the g.i. problem in the e.m. contribution
to the $n-p$ mass difference.

\section*{Appendix A: Evaluation of the Integral (4.5)}
\setcounter{equation}{0}
\renewcommand{\theequation}{A.\arabic{equation}}
\par
        The master expression (4.8) after being substituted in the full e.m.
 self energy contribution (4.5) is integrated over each $d\sigma_{12}$ and
 $d{\sigma_{12}}'$. The final result is
\begin{eqnarray}
\delta M^{\gamma} & = & \sum_{123} {{2e^2} \over {9}} \tau_3 \int
{(m_q + \omega_3)\over {2\omega_3}} {d^3{\hat p}_3\over {4\pi}} {d^3{\hat
q}_{12}\over {4\pi}}
{d^3{{\hat q}_{12}}'\over {4\pi}} {1 \over {2{\hat q}_{12} 2{{\hat
q}_{12}}'}} \times \nonumber \\
                  &   & F({\hat q}_{12}, {{\hat q}_{12}}', {\hat p}_3 )
exp{(-{2\over 3}[{{\hat q}_{12}}^2 + {{\hat q}_{12}}^{'2} + 3{{\hat
p}_3}^2]/{\beta}^2)}
\end{eqnarray}
where
\begin{eqnarray}
\lefteqn{F({\hat q}_{12}, {{\hat q}_{12}}', {\hat p}_3 ) = }  \nonumber \\
  & & (m_q + \omega_{12})(m_q + {\omega_{12}}') k^2 + (M\omega_3 - {1\over
2}
(M^2 - m_q^2))[(m_q + \omega_{12})^2 \nonumber \\
  & & + (m_q + {\omega_{12}}')^2 + (m_q + \omega_{12})(m_q + {\omega_{12}}')
- (m_q + \omega_{12})^2 {D_{12}}'/{2{\omega_{12}}'} \nonumber \\
  & & - (m_q + {\omega_{12}}')^2 D_{12}/{2\omega_{12}}
- (m_q + \omega_{12})(m_q + {\omega_{12}}')[D_{12}/{2\omega_{12}} \nonumber
\\
  & & + {D_{12}}'/{2{\omega_{12}}'}] k^2 [(m_q + 2 \omega_{12})(m_q
+ 2{\omega_{12}}') - m_q^2]/2 \nonumber \\
  & & - [(m_q + 2 \omega_{12})(m_q + 2{\omega_{12}}') - m_q^2]
[D_{12}/{2\omega_{12}} + {D_{12}}'/{2{\omega_{12}}'}]/2  \nonumber \\
  & & -{1\over 8} {D_{12} {D_{12}}'\over {\omega_{12} {\omega_{12}}'}} + k^4
-  k^2 [D_{12}/{\omega_{12}} + {D_{12}}'/{\omega_{12}}']
\end{eqnarray}
Using eq.(4.9), the integration over ${\hat q}_{12}$  and ${{\hat q}_{12}}'$
can be done independently of each other, and thus can be written in a
compact
notation as follows
\begin{equation}
\delta M^{\gamma} = \sum_{123} (+{2\over 9}e^2 \tau_3) \int {\hat p}_3^2
d{\hat p}_3 {(m_q + \omega_3)\over {2\omega_3}} F_1 \ e^{[- {\hat
p}_3^2/\beta^2]}
\end{equation}
where
\begin{eqnarray}
F_1 & = & J_{11}J_{11} + [M\omega_3 -{1\over 2}M^2 +{1\over 2}m_q^2]
(2J_{20}J_{00} + 2J_{10}J_{10}) - 2J_{20}I_{00} \nonumber \\
    &   & - 2J_{10}I_{10} + {J_{11}}'{J_{11}}'/2 -J_{01}J_{01}m_q^2/2
- {I_{10}}'{J_{10}}' \nonumber \\
    &   & + m_q^2 I_{00}J_{00} - I_{00}I_{00}/2 -J_{02}J_{02}/4 +
I_{01}I_{01}
\end{eqnarray}
and with $(n = 0,1,2 ; m = 0,1,2)$,
\begin{equation}
J_{nm};I_{nm} = 2^{-1/2} \int {\hat q}_{12} d{\hat q}_{12} \ e^{(-{2\over 3}
{{\hat q}_{12}}^2/\beta^2)}[{\sqrt 2}{\hat q}_{12}]^m (m_q + \omega_{12})^n
[1 ;  {1\over 2}D_{12}/{\omega_{12}}];
\end{equation}

\begin{equation}
{J_{nm}}';{I_{nm}}' = 2^{-1/2} \int {\hat q}_{12} d{\hat q}_{12} \
e^{(-{2\over 3}
{{\hat q}_{12}}^2/\beta^2)} [{\sqrt 2}{\hat q}_{12}]^m (m_q + 2
\omega_{12})^n
[1 ;  {1\over 2}D_{12}/{\omega_{12}}];
\end{equation}

\section*{Appendix B: Derivation of qqq Vertex Fn, Eq.(2.5)}
\subsection*{B.1: Method of Green's Functions}
\setcounter{equation}{0}
\renewcommand{\theequation}{B.1.\arabic{equation}}
\par
        We outline here some essential steps leading to a formal derivation
of
eq.(2.5) which was written down in a semi-intuitive fashion in [12]. To that
end we shall employ the method of Green's functions for 2- and 3- particle
scattering {\it {near the bound state pole}}, since the inhomogeneous terms
are not relevant for our purposes. For simplicity we shall consider
identical
{\it spinless} bosons, with pairwise BS kernels under CIA conditions [8],
first for the 2-body case for calibration, and then for the 3-body system.
\subsection*{B.2: Two-Quark Green's Function}
\setcounter{equation}{0}
\renewcommand{\theequation}{B.2.\arabic{equation}}
\par
        Apart from some results already goven in the text, we shall use the
notation and phase conventions of [8,12] for the various quantities
(momenta,
propagators, etc). The 4D $qq$ Green's function $G(p_1p_2 ; {p_1}'{p_2}')$
near a {\it bound} state satisfies a 4D BSE without the inhomogeneous term,
viz. [8,12],
\begin{equation}
i(2\pi)^4 G(p_1 p_2;{p_1}'{p_2}') = {\Delta_1}^{-1} {\Delta_2}^{-1} \int
d{p_1}'' d{p_2}'' K(p_1 p_2;{p_1}''{p_2}'') G({p_1}''{p_2}'';{p_1}'{p_2}')
\end{equation}
 where
\begin{equation}
\Delta_1 = {p_1}^2 + {m_q}^2 ,
\end{equation}
and $m_q$ is the mass of each quark. Now using the relative 4- momentum
$q = (p_1-p_2)/2$ and total 4-momentum $P = p_1 + p_2$
(similarly for the other sets), and removing a $\delta$-function
for overall 4-momentum conservation, from each of the $G$- and $K$-
functions, eq.(B.2.1) reduces to the simpler form
\begin{equation}
i(2\pi)^4 G(q.q') = {\Delta_1}^{-1} {\Delta_2}^{-1}  \int d{\hat q}''
Md{\sigma}'' K({\hat q},{\hat q''}) G(q'',q')
\end{equation}
where ${\hat q}_{\mu} = q_{\mu} - {\sigma} P_{\mu}$, with
$\sigma = (q.P)/P^2$, is effectively 3D in content (being orthogonal to
$P_{\mu}$). Here we have incorporated the ansatz of a 3D support for the
kernel $K$ (independent of $\sigma$ and ${\sigma}'$), and broken up the
4D measure $dq''$ arising from (2.1) into the product
$d{\hat q}''Md{\sigma}''$ of a 3D and a 1D measure respectively. We have
also suppressed the 4-momentum $P_{\mu}$ label, with $(P^2 = -M^2)$, in
the notation for $G(q.q')$.
\par
        Now define the fully 3D Green's function
${\hat G}({\hat q},{\hat q}')$ as [29]
\begin{equation}
{\hat G}({\hat q},{\hat q}') = \int \int M^2 d{\sigma}d{\sigma}'G(q,q')
\end{equation}
and two (hybrid) 3D-4D Green's functions ${\tilde G}({\hat q},q')$,
${\tilde G}(q,{\hat q}')$ as
\begin{equation}
{\tilde G}({\hat q},q') = \int Md{\sigma} G(q,q');
{\tilde G}(q,{\hat q}') = \int Md{\sigma}' G(q,q');
\end{equation}
Next, use (B.2.5) in (B.2.3) to give
\begin{equation}
i(2\pi)^4 {\tilde G}(q,{\hat q}') = {\Delta_1}^{-1} {\Delta_2}^{-1}
\int dq'' K({\hat q},{\hat q}''){\tilde G}(q'',{\hat q}')
\end{equation}
Now integrate both sides of (B.2.3) w.r.t. $Md{\sigma}$ and use the result
[8]
\begin{equation}
\int Md{\sigma}{\Delta_1}^{-1} {\Delta_2}^{-1} = 2{\pi}i D^{-1}({\hat q});
\quad D({\hat q}) = 4{\hat \omega}({\hat \omega}^2 - M^2/4);\quad
{\hat \omega}^2 = {m_q}^2 + {\hat q}^2
\end{equation}
to give a 3D BSE w.r.t. the variable ${\hat q}$, while keeping the other
variable $q'$ in a 4D form:
\begin{equation}
(2\pi)^3 {\tilde G}({\hat q},q') = D^{-1} \int d{\hat q}''
K({\hat q},{\hat q}'') {\tilde G}({\hat q}'',q')
\end{equation}
Now a comparison of (B.2.3) with (B.2.8) gives the desired connection
between
the full 4D $G$-function and the hybrid ${\tilde G({\hat q}, q')}$-function:
\begin{equation}
2{\pi}i G(q,q') = D({\hat q}){\Delta_1}^{-1}{\Delta_2}^{-1}
{\tilde G}({\hat q},q')
\end{equation}
Again, the symmetry of the left hand side of (B.2.9) w.r.t. $q$ and $q'$
allows us to write the right hand side with the roles of $q$ and $q'$
interchanged. This gives the dual form
\begin{equation}
2{\pi}i G(q,q') = D({\hat q}'){{\Delta_1}'}^{-1}{{\Delta_2}'}^{-1}
{\tilde G}(q,{\hat q}')
\end{equation}
which on integrating both sides w.r.t. $M d{\sigma}$ gives
\begin{equation}
2{\pi}i{\tilde G}({\hat q},q') = D({\hat q}'){{\Delta_1}'}^{-1}
{{\Delta_2}'}^{-1}{\hat G}({\hat q},{\hat q}').
\end{equation}
Substitution of (B.2.11) in (B.2.9) then gives the symmetrical form
\begin{equation}
(2{\pi}i)^2 G(q,q') = D({\hat q}){\Delta_1}^{-1}{\Delta_2}^{-1}
{\hat G}({\hat q},{\hat q}')D({\hat q}'){{\Delta_1}'}^{-1}
{{\Delta_2}'}^{-1}
\end{equation}
Finally, integrating both sides of (B.2.8) w.r.t. $M d{\sigma}'$, we
obtain a fully reduced 3D BSE for the 3D Green's function:
\begin{equation}
(2\pi)^3 {\hat G}({\hat q},{\hat q}') = D^{-1}({\hat q} \int d{\hat q}''
K({\hat q},{\hat q}'') {\hat G}({\hat q}'',{\hat q}')
\end{equation}
Eq.(B.2.12) which is valid near the bound state pole (since the
inhomogeneous term has been dropped for simplicity) expresses the desired
connection between the 3D and 4D forms of the Green's functions; and
eq(B.2.13) is the determining equation for the 3D form. A spectral analysis
can now be made for either of the 3D or 4D Green's functions in the
standard manner, viz.,
\begin{equation}
G(q,q') = \sum_n {\Phi}_n(q;P){\Phi}_n^*(q';P)/(P^2 + M^2)
\end{equation}
where $\Phi$ is the 4D BS wave function. A similar expansion holds for
the 3D $G$-function ${\hat G}$ in terms of ${\phi}_n({\hat q})$.
Substituting
these expansions in (B.2.12), one immediately sees the connection between
the 3D and 4D wave functions in the form:
\begin{equation}
2{\pi}i{\Phi}(q,P) = {\Delta_1}^{-1}{\Delta_2}^{-1}D(\hat q){\phi}(\hat q)
\end{equation}
whence the BS vertex function becomes $\Gamma = D \times \phi/(2{\pi}i)$
as found in [8]. We shall make free use of these results, taken as $qq$
subsystems, for our study of the $qqq$ $G$-functions in Sections 3 and 4.

\subsection*{B.3: 3D Reduction of the BSE for 3-Quark G-function}
\setcounter{equation}{0}
\renewcommand{\theequation}{B.3.\arabic{equation}}
\par
        As in the two-body case, and in an obvious notation for various
4-momenta (without the Greek suffixes), we consider the most general
Green's function $G(p_1 p_2 p_3;{p_1}' {p_2}' {p_3}')$ for 3-quark
scattering {\it near the bound state pole} (for simplicity) which allows
us to drop the various inhomogeneous terms from the beginning. Again we
take out an overall delta function $\delta(p_1 + p_2 + p_3 - P)$ from the
$G$-function  and work with {\it two} internal 4-momenta for each of the
initial and final states defined as follows [12]:
\begin{equation}
{\sqrt 3}{\xi}_3 =p_1 - p_2 \ ; \quad  3{\eta}_3 = - 2p_3 + p_1 +p_2
\end{equation}
\begin{equation}
P = p_1 + p_2 + p_3 = {p_1}' + {p_2}' + {p_3}'
\end{equation}
and two other sets ${\xi}_1,{\eta}_1$ and ${\xi}_2,{\eta}_2$ defined by
cyclic permutations from (B.3.1). Further, as we shall consider pairwise
kernels with 3D support, we define the effectively 3D momenta ${\hat p}_i$,
as well as the three (cyclic) sets of internal momenta
${\hat \xi}_i,{\hat \eta}_i$, (i = 1,2,3) by [12]:
\begin{equation}
{\hat p}_i = p_i - {\nu}_i P \ ;\quad  {\hat {\xi}}_i = {\xi}_i - s_i P\  ;
\quad
{\hat {\eta}}_i - t_i P
\end{equation}
\begin{equation}
{\nu}_i = (P.p_i)/P^2\  ;\quad s_i = (P.\xi_i)/P^2 \ ;\quad t_i =
(P.\eta_i)/P^2 \end{equation}
\begin{equation}
{\sqrt 3} s_3 = \nu_1 - \nu_2 \ ;\quad 3 t_3 = -2 \nu_3 + \nu_1 + \nu_2 \
\ ( + {\rm cyclic permutations})
\end{equation}
The space-like momenta ${\hat p}_i$ and the time-like ones $\nu_i$
satisfy [12]
\begin{equation}
{\hat p}_1 + {\hat p}_2 + {\hat p}_3 = 0\  ;\quad \nu_1 + \nu_2 + \nu_3 = 1
\end{equation}
Strictly speaking, in the spirit of covariant instantaneity, we should
have taken the relative 3D momenta ${\hat \xi},{\hat \eta}$ to be in the
instantaneous frames of the concerned pairs, i.e., w.r.t. the rest frames
of $P_{ij} = p_i +p_j$; however the difference between the rest frames of
$P$ and $P_{ij}$  is small and calculable [12], while the use of a common
3-body rest frame $(P = 0)$ lends considerable simplicity and elegance to
the formalism.
\par
        We may now use the foregoing considerations to write down the BSE
for the 6-point Green's function in terms of relative momenta, on closely
parallel lines to the 2-body case. To that end note that the 2-body
relative momenta are $q_{ij} = (p_i - p_j)/2 = {\sqrt 3}{\xi_k}/2$, where
(ijk) are cyclic permutations of (123). Then for the reduced $qqq$ Green's
function, when the {\it last} interaction was in the (ij) pair, we may use
the notation $G(\xi_k \eta_k ; {\xi_k}' {\eta_k}')$, together with `hat'
notations on these 4-momenta when the corresponding time-like components
are integrated out. Further, since the pair $\xi_k,\eta_k$ is
{\it {permutation invariant}} as a whole, we may choose to drop the index
notation from the complete $G$-function to emphasize this symmetry as and
when needed. The $G$-function for the $qqq$ system satisfies, in the
neighbourhood of the bound state pole, the following (homogeneous) 4D BSE
for pairwise $qq$ kernels with 3D support:
\begin{equation}
i(2\pi)^4 G(\xi \eta ;{\xi}' {\eta}') = \sum_{123}
{\Delta_1}^{-1} {\Delta_2}^{-1} \int d{{\hat q}_{12}}'' M d{\sigma_{12}}''
K({\hat q}_{12}, {{\hat q}_{12}}'') G({\xi_3}'' {\eta_3}'';{\xi_3}'
{\eta_3}')
\end{equation}
where we have employed a mixed notation ($q_{12}$ versus $\xi_3$) to stress
the two-body nature of the interaction with one spectator at a time, in a
normalization directly comparable with eq.(B.2.3) for the corresponding
two-body problem. Note also the connections
\begin{equation}
\sigma_{12} = {\sqrt 3}{s_3}/2   ;\quad
{\hat q}_{12} = {\sqrt 3}{{\hat \xi}_3}/2  ; \quad {\hat \eta}_3 =
-{\hat p}_3, \quad etc
\end{equation}
The next task is to reduce the 4D BSE (B.3.7) to a fully 3D form through a
sequence of integrations w.r.t. the time-like momenta $s_i,t_i$ applied
to the different terms on the right hand side, {\it {provided both}}
variables are simultaneously permuted. We now define the following fully
3D as well as mixed (hybrid) 3D-4D $G$-functions according as one or more
of the time-like $\xi,\eta$ variables are integrated out:
\begin{equation}
{\hat G}({\hat \xi} {\hat \eta};{\hat \xi}' {\hat \eta}') =
\int \int \int \int ds dt ds' dt' G(\xi \eta ; {\xi}' {\eta}')
\end{equation}
which is $S_3$-symmetric.
\begin{equation}
{\tilde G}_{3\eta}(\xi {\hat \eta};{\xi}' {\hat \eta}') =
\int \int dt_3 d{t_3}' G(\xi \eta ; {\xi}' {\eta}');
\end{equation}
\begin{equation}
{\tilde G}_{3\xi}({\hat \xi}  \eta;{\hat \xi}' {\eta}') =
\int \int ds_3 d{s_3}' G(\xi \eta ; {\xi}' {\eta}');
\end{equation}
The last two equations are however {\it not} symmetric w.r.t. the
permutation group $S_3$, since both the variables ${\xi,\eta}$ are not
simultaneously transformed; this fact has been indicated in eqs.(B.3.10-11)
by the suffix ``3" on the corresponding (hybrid) ${\tilde G}$-functions,
to emphasize that the `asymmetry' is w.r.t. the index ``3". We shall term
such quantities ``$S_3$-indexed", to distinguish them from $S_3$-symmetric
quantities as in eq.(B.3.9). The full 3D BSE for the ${\hat G}$-function is
obtained by integrating out both sides of (B.3.7) w.r.t. the $st$-pair
variables $ds_i d{s_j}' dt_i d{t_j}'$ (giving rise to an $S_3$-symmetric
quantity), and using (B.3.9) together with (B.3.8) as follows:
\begin{equation}
(2\pi)^3 {\hat G}({\hat \xi} {\hat \eta} ;{\hat \xi}' {\hat \eta}') =
\sum_{123} D^{-1}({\hat q}_{12}) \int d{{\hat q}_{12}}''
K({\hat q}_{12}, {{\hat q}_{12}}'') {\hat G}({\hat \xi}'' {\hat \eta}'';
{\hat \xi}' {\hat \eta}')
\end{equation}
This integral equation for ${\hat G}$ which is the 3-body counterpart of
(B.2.13) for a $qq$ system in the neighbourhood of the bound state pole,
is the desired 3D BSE for the $qqq$ system in a {\it {fully connected}}
form, i.e., free from delta functions. Now using a spectral decomposition
for ${\hat G}$
\begin{equation}
{\hat G}({\hat \xi} {\hat \eta};{\hat \xi}' {\hat \eta}')
= \sum_n {\phi}_n( {\hat \xi} {\hat \eta} ;P)
{\phi}_n^*({\hat \xi}' {\hat \eta}';P)/(P^2 + M^2)
\end{equation}
on both sides of (B.3.12) and equating the residues near a given pole
$P^2 = -M^2$, gives the desired equation for the 3D wave function $\phi$
for the bound state in the connected form:
\begin{equation}
(2\pi)^3 \phi({\hat \xi} {\hat \eta} ;P) = \sum_{123} D^{-1}({\hat q}_{12})
\int d{{\hat q}_{12}}'' K({\hat q}_{12}, {{\hat q}_{12}}'')
\phi({\hat \xi}'' {\hat \eta}'' ;P)
\end{equation}
Now the $S_3$-symmetry of $\phi$ in the $({\hat \xi}_i, {\hat \eta}_i)$ pair
is a very useful result for both the solution of (B.3.14) {\it and} for the
reconstruction of the 4D BS wave function in terms of the 3D wave function
(B.3.14), as is done in the subsection below.
\subsection*{B.4: Reconstruction of the 4D BS Wave Function}
\setcounter{equation}{0}
\renewcommand{\theequation}{B.4.\arabic{equation}}
\par
        We now attempt to {\it re-express} the 4D $G$-function given by
(B.3.7) in terms of the 3D ${\hat G}$-function given by (B.3.12), as the
$qqq$ counterpart of the $qq$ results (B.2.12-13). To that end we adapt
the result (B.2.12) to the hybrid Green's function  of the (12) subsystem
given by ${\tilde G}_{3 \eta}$, eq.(B.3.10), in which the 3-momenta
${\hat \eta}_3,{{\hat \eta}_3}'$ play a parametric role reflecting the
spectator status of quark $\# 3$, while the {\it active} roles are played
by $q_{12}, {q_{12}}' = {\sqrt 3}(\xi_3,{\xi_3}')/2$, for which the analysis
of subsec.B.2 applies directly. This gives
\begin{equation}
(2{\pi}i)^2 {\tilde G}_{3 \eta}(\xi_3 {\hat \eta}_3;
{\xi_3}' {{\hat \eta}_3}')
= D({\hat q}_{12}){\Delta_1}^{-1}{\Delta_2}^{-1}
{\hat G}({\hat \xi_3} {\hat \eta_3}; {\hat \xi_3}' {\hat \eta_3}')
D({{\hat q}_{12}}'){{\Delta_1}'}^{-1}{{\Delta_2}'}^{-1}
\end{equation}
where on the right hand side, the `hatted' $G$-function has full
$S_3$-symmetry, although (for purposes of book-keeping) we have not
shown this fact explicitly by deleting the suffix `3' from its
arguments. A second relation of this kind may be obtained from (B.3.7)
by noting that the 3 terms on its right hand side may be expressed in
terms of the hybrid ${\tilde G}_{3 \xi}$ functions vide their definitions
(B.3.11), together with the 2-body interconnection between
$(\xi_3,{\xi_3}')$
and $({\hat \xi}_3,{{\hat \xi}_3}')$ expressed once again via (B.4.1), but
without the `hats' on $\eta_3$ and ${\eta_3}'$. This gives
\begin{eqnarray}
({\sqrt 3} \pi i)^2 G(\xi_3 \eta_3; {\xi_3}'{\eta_3}')
&=& ({\sqrt 3} \pi i)^2 G(\xi \eta; {\xi}'{\eta}')\nonumber\\
&=& \sum_{123} {\Delta_1}^{-1}{\Delta_2}^{-1} (\pi i {\sqrt 3})
\int d{{\hat q}_{12}}'' M d{\sigma_{12}}''
K({\hat q}_{12}, {{\hat q}_{12}}'')
G({\xi_3}'' {\eta_3}'';{\xi_3}' {\eta_3}')\nonumber\\
&=& \sum_{123} D({\hat q}_{12}) {\Delta_1}^{-1}{\Delta_2}^{-1}
{\tilde G}_{3 \xi}({\hat \xi}_3  \eta_3; {{\hat \xi}_3}' {{\eta}_3}')
{{\Delta_1}'}^{-1} {{\Delta_2}'}^{-1}
\end{eqnarray}
where the second form exploits the symmetry between $\xi,\eta$ and
$\xi',\eta'$.
\par
        At this stage, unlike the 2-body case, the reconstruction of the
4D Green's function is {\it {not yet}} complete for the 3-body case, as
eq.(B.4.2) clearly shows. This is due to the {\it truncation} of Hilbert
space implied in the ansatz of 3D support to the pairwise BSE kernel $K$
which, while facilitating a 4D to 3D BSE reduction without extra charge,
does {\it not} have the {\it complete} information to permit the {\it
reverse}
transition (3D to 4D) without additional assumptions; see [29] for details.
The physical reasons for the 3D ansatz for the BSE kernel have been
discussed
in detail elsewhere [23,29], vis-a-vis contemporary approaches. Here we look
upon this ``inverse" problem as a purely {\it mathematical} one.
\par
        We must now look for a suitable  ansatz for the quantity
${\tilde G}_{3 \xi}$ on the right hand side of (B.4.2) in terms of {\it
known}
quantities, so that the reconstructed 4D $G$-function satisfies the 3D
equation (B.3.12) exactly, as a ``check-point" for the entire exercise. We
therefore seek a structure of the form
\begin{equation}
{\tilde G}_{3 \xi}({\hat \xi}_3  {\eta}_3; {{\hat \xi}_3}' {{\eta}_3}')
= {\hat G}({{\hat \xi}_3} {\hat \eta}_3; {{\hat \xi}_3}' {{\hat \eta}_3}')
\times F(p_3, {p_3}')
\end{equation}
where the unknown function $F$ must involve only the momentum of the
spectator quark $\# 3$. A part of the $\eta_3, {\eta_3}'$ dependence has
been absorbed in the ${\hat G}$ function on the right, so as to satisfy
the requirements of $S_3$-symmetry for this 3D quantity [29].
\par
        As to the remaining factor $F$, it is necessary to choose its
form in a careful manner so as to conform to the conservation of
4-momentum for the {\it free} propagation of the spectator between two
neighbouring vertices, consistently with the symmetry between $p_3$
and ${p_3}'$. A possible choice consistent with these conditions is
the form (see [29] for details):
\begin{equation}
F(p_3, {p_3}') = C_3 {\Delta_3}^{-1} {\delta}(\nu_3 - {\nu_3}')
\end{equation}
Here ${\Delta_3}^{-1}$ represents the ``free" propagation of quark $\# 3$
between successive vertices, while $C_3$ represents some residual effects
which may at most depend on the 3-momentum ${\hat p}_3$, but must satisfy
the main constraint that the 3D BSE, (B.3.12), be {\it explicitly}
satisfied.
\par
        To check the self-consistency of the ansatz (B.4.4), integrate
both sides of (B.4.2) w.r.t. $ds_3 d{s_3}' dt_3 d{t_3}'$ to recover the
3D $S_3$-invariant ${\hat G}$-function on the left hand side. Next, in
the first form on the right hand side, integrate w.r.t. $ds_3 d{s_3}'$
on the $G$-function which alone involves these variables. This yields
the quantity ${\tilde G}_{3 \xi}$. At this stage, employ the ansatz
(B.4.4) to integrate over $dt_3 d{t_3}'$. Consistency with the 3D BSE,
eq.(B.3.12), now demands
\begin{equation}
C_3 \int \int d\nu_3 d{\nu_3}' {\Delta_3}^{-1} \delta(\nu_3 - {\nu_3}')
= 1 ; (since dt = d\nu)
\end{equation}
The 1D integration w.r.t. $d\nu_3$ may be evaluated as a contour
integral over the propagator ${\Delta}^{-1}$ , which gives the pole
at $\nu_3 = {\hat \omega}_3/M$, (see below for its definition). Evaluating
the residue then gives
\begin{equation}
C_3 = i \pi / (M {\hat \omega}_3 ) ;  \quad
{{\hat \omega}_3}^2 = {m_q}^2 + {{\hat p}_3}^2
\end{equation}
which will reproduce the 3D BSE, eq.(B.3.12), {\it exactly}! Substitution
of (B.4.4) in the second form of (B.4.2) finally gives the desired 3-body
generalization of (B.2.12) in the form
\begin{equation}
3 G(\xi \eta; \xi' \eta') = \sum_{123} D({\hat q}_{12}) \Delta_{1F}
\Delta_{2F} D({{\hat q}_{12}}') {\Delta_{1F}}' {\Delta_{2F}}'
{\hat G}({\hat \xi_3} {\hat \eta_3}; {\hat \xi_3}' {\hat \eta_3}')
[\Delta_{3F} / (M \pi {\hat \omega}_3)]
\end{equation}
where for each index, $\Delta_F = - i {\Delta}^{-1}$ is the
Feynman propagator.
\par
        To find the effect of the ansatz (B.4.4) on the 4D BS
{\it {wave function}} $\Phi(\xi \eta; P)$, we do a spectral reduction
like (B.3.13) for the 4D Green's function $G$ on the left hand side of
(B.4.2). Equating the residues on both sides gives the desired 4D-3D
connection between $\Phi$ and $\phi$:
\begin{equation}
\Phi(\xi \eta; P) = \sum_{123} D({\hat
q}_{12}){\Delta_1}^{-1}{\Delta_2}^{-1}
\phi ({\hat \xi} {\hat \eta}; P) \times
\sqrt{{\delta(\nu_3 -{\hat \omega}_3/M)} \over{M {\hat \omega}_3
{\Delta}_3}}
\end{equation}
>From (B.4.8) and eq.(2.1) of the text, we infer the structure of the
baryon-$qqq$ vertex function $V_3$ as given in eq.(2.5) of the text. For
a detailed discussion of the significance of this result, vis-a-vis
contemporary approaches, see [29].

\section*{Appendix C: Gauge Corrections to Kaon E.M. Mass}
\setcounter{equation}{0}
\renewcommand{\theequation}{C.\arabic{equation}}
\par
        We outline here a practical procedure to evaluate the gauge
corrections
to the e.m. self-energy of a $q{\bar q}$ system, vide fig.1b of [28],
pending
a more systematic treatment in a later paper. This two-body exercise
should hopefully serve as a fascimile of the effect expected for the present
$qqq$ case. For brevity we shall refer to the figures of KL [32] in their
notation without drawing them anew. Thus fig 1b of [28] corresponds
to fig 1a of KL [32], except for the presence of the hadron lines at the two
ends. We shall call this simply `1a', with the understanding that the hadron
lines are `attached' to 1a.  For the actual mathematical symbols (including
phase conventions) we shall draw freely from [28], without explanation. In
[28], only 1a of [32] was calculated, but now one must add 2(a,b,c,d,e) of
[32], all with hadron lines understood at the two ends of each. There is no
need to calculate 1b or 1c of [32] which are mere e.m. self-energies of
single
quarks (g.i. by themselves), and are routinely absorbed in quark mass
renormalization (of little significance in this phenomenological study which
has these masses as inputs).
\par
        A new ingredient is a 4-point vertex in each of 2(a,b,c,d), and
{\it two} 4-point vertices in 2e, except that the word `point' is now
understood as an extended structure characterized by the hadron-quark vertex
function $D({\hat q})\phi({\hat q})$ where one must insert a photon line in
each such $Hq{\bar q}$ blob. Since it is {\it not} a standard point vertex,
the method [32] of inserting exponential phase integrals with each current
is
not technically feasible; instead we may resort to the simple-minded
substition $p_i-e_iA(x_i)$ for each 4-momentum $p_i$ (in a mixed $p,x$
representation) occurring in the structure of the vertex function, which has
the same physical content, at least up to first order in the e.m. field,
without further comment. This amounts to replacing each ${\hat q}_\mu$
occurring in $\Gamma({\hat q})$ = $D({\hat q})\phi({\hat q})$, by
${\hat q}_\mu -e_q A_\mu$, where $e_q = {\hat m}_2 e_1 -{\hat m}_1 e_2$.
The net result in the first order in $A_\mu$ is a first order correction
to $\Gamma({\hat q})$ of amount $e_q j({\hat q}).A$ defined by
\begin{equation}
j({\hat q}).A = -4M{\hat q}.A\phi({\hat q}) (1-D({\hat q})/(4M \beta^2))
\end{equation}
where we have made free use of various symbols and definitions in [28].
(The effect of the hat structure of ${\hat q}$ on the e.m. substitution
is ignored in this approximate treatment). This effective 4-point vertex
function is operative at one end in each of 2a,2b,2c,2d of KL [32] and
at both ends of 2e. For the e.m. vertex at the quark lines of 2(a,b,c,d),
we use simply $ie_i \gamma.A$, as in [28]. The matrix elements can now be
written down on exactly the same lines, and the {\it same} phase convention
as in [28] to keep proper track of the gauge corrections with sign. We need
write these down only for 2a and 2e, noting the equalities 2a=2b, as also
2c=2d, and the further substitutions $(1) \rightarrow (2)$ and vice versa
to generate 2c(=2d) from 2a(=2b). The contribution from 2a [32] to the e.m.
quadratic self-energy of a kaon is expressible as
\begin{eqnarray}
M^2_{2a} &=& N_H^2 (2\pi)^{-5}e_1 e_q \int  j({\hat q})_\mu D({\hat q}')
\phi({\hat q}')Tr[\gamma_5 D_{F\mu \nu}(k) \nonumber \\
         & & S_F(p_1-{\hat m}_1 k) i e_1 \gamma_\nu
S_F(p_1') \gamma_5 S_F(-p_2')] d^4q d^4k
\end{eqnarray}
where $p_1'=p_1+{\hat m}_2 k$ and $p_2 = p_2' = p_2-{\hat m}_2 k$ are the
4-momenta of the quarks at the other (right-hand) end, and the photon
propagator in the Landau gauge is $-i(\delta_{\mu\nu} -k_\mu
k_\nu/k^2)/k^2$.
To make better use of the techniques outlined in [28], it is convenient to
change the variable from $k_\mu$ to $q_\mu'$, noting that
$q'= q+{\hat m}_2 k$, which gives  $d^4 k = d^4 q'/{\hat m}_2^4$, etc.
This shows that fig 2a(=2b), where the photon line ends on the heavier quark
$m_1$, gives a bigger contribution than does fig.2c(=2d) which would give
${\hat m}_1^{-4}$ arising from the $d^4k$-measure. Evaluating the traces,
and
integrating over the poles of the two time-like momenta $q_0$ and $q_0'$
gives for the sum of the contributions from 2a-2d to the quadratic mass
difference between ${\bar K}^0$ and $K^-$ as a product of two 3D quadratures
after some simplifications with factorable approximations a la [28]:
\begin{eqnarray}
\delta M^2_{2(a-d)} &=& {{6N_H^2 M \delta (e_1 e_q)}\over {(2\pi)^3
{\hat m}_2^3}} \int d^3{\hat q} \int d^3{\hat q}'{{\phi \phi'}\over
{{\hat q}{\hat q}'\omega_{1k}}}[1- {{D({\hat q})}\over{4M\beta^2}}]\nonumber
\\
                    & & [({\hat q}^2 (2-4/\pi)-{\hat q}{\hat q}'/3)
(M^2-{\delta m}^2+D({\hat q}'){\omega_1'}^{-1}/2+
D({\hat q}'){\omega_2'}^{-1}/2)\nonumber \\
                    & & + {1 \over 3}{\hat m}_2{\hat q}{\hat q}'
(D({\hat q}'){\omega_2'}^{-1}/2+ M^2-{\delta m}^2)] + [1 \leftrightarrow 2]
\end{eqnarray}
Here $\delta (e_i e_q)$ is the ${\bar K}^0$ minus $K^-$ difference
between the indicated charge factors associated with line `i', while
$\omega_{1,2}'^2 = m_{1,2}^2 + {\hat q}'^2$ and
$\omega_{1k}^2 = m_1^2 + ({\hat q}-{\hat m}_1 {\hat k})^2$.
\par
        Next the contribution to $\delta M^2$ arising from fig 2e of KL [32]
which involves the product of two vertex blobs like (C.1) is given by
\begin{equation}
\delta M^2_{2e} = i N_H^2 (2\pi)^{-5}{e_q}^2 \int d^4q d^4k D_{F\mu\nu}(k)
j({\hat q})_\mu j({\hat q})_\nu
Tr[\gamma_5 S_F(p_1-{\hat m}_1 k) \gamma_5 S_F(-p_2+{\hat m}_2 k)]
\end{equation}
This integral is somewhat different in structure from (C.2) in as much as
$k_\mu$ is fully decoupled from either wave function $\phi, \phi'$, both of
which have the same argument ${\hat q}$. This makes it possible to integrate
first over $d^4k$ as well as the time-like component $q_0$ of $q_\mu$
neither of which is involved in the vertex function. The relevant integral
after tracing and rearranging has the form
\begin{eqnarray}
F({\hat q}) &=& 3(-i)^2\int d^4k \int dq_0 k^{-2}
(\delta_{\mu\nu}-k_\mu k_\nu/k^2) \nonumber \\
            & & [{\hat q}^2 -q_0^2 + m_1m_2
-{\hat m}_1{\hat m}_2 (P-k)^2]/(\Delta_1\Delta_2)
\end{eqnarray}
where $\Delta_i = m_i^2 + (p_i - {\hat m}_i k)^2$. The integral which is
entirely convergent works out after some standard manipulations involving
Feynman techniques as well as differentiation under integral signs as
\begin{equation}
F({\hat q})= 6\pi^3 [m_1m_2+{\hat q}^2+\Lambda][\sqrt{\Lambda}-
\sqrt{\Lambda-{\hat m}_1{\hat m}_2 M^2}]/({\hat m}_1{\hat m}_2 M)^2
\end{equation}
where $\Lambda= {\hat m}_1{\hat m}_2 M^2 + D({\hat q})/2M$. And the final
expression for (C.4) in terms of (C.6) is
\begin{equation}
\delta M^2_{2e} = N_H^2 (2\pi)^{-5} {\delta (e_q^2)} \int d^3{\hat q}
{j({\hat q})}^2 F({\hat q})
\end{equation}
\par
        Further evaluation of (C.3) and (C.7) can be made a la [28] in a
straightforward way. The key ingredients are
\begin{equation}
{\delta {e_1e_q}}= 0.236 e^2 ; \quad {\delta {e_2e_q}}=0.139 e^2;
\quad {\delta {e_q^2}}= -0.0294 e^2.
\end{equation}
The break-up of the final results for the diagrams 2(a-e) after dividing the
results of(C.3) and (C.7) by 2M, since $\delta M^2 = 2M \delta M$, is (in
MeV):
\begin{equation}
\delta M_{2a+2b}=-0.6996; \quad \delta M_{2c+2d}=+0.1358; \quad
\delta M_{2e}=-0.0481; \quad \delta M_{tot}=-0.612 MeV.
\end{equation}
All these corrections, which reinforce one another due to a complex
interplay
of signs, add up to a figure which increases the value -1.032 MeV due to
fig 1(b) found in [28], to -1.644 MeV.  A more comprehensive paper for
the gauge corrections to fig 1(a) of [32] for the other mesons, as well as
the
$qqq$ baryon, will be communicated separately. This estimate has been used,
on a percentage basis, as a facsimile for the gauge correction expected
for the e.m. self energy of the nucleon, in sec.5.3 of text.

\section*{Figure Captions}

\par
Fig.1: Diagrams for BS normalization of Baryon-$qqq$ vertex function. 1(a)
shows quark $\# 1$ emitting a zero momentum photon ($k=0$); its last $qq$
interaction was with $\# 2$, while $\# 3$ is the spectator. 1(b) is the
same diagram with the roles of $\# 1$ and $\# 2$ interchanged. 1(c)
denotes schematically two more such pairs of diagrams obtained with cyclical
permutations of the indices (123) in pairs. The 4-momenta on the quark lines
are shown as used in the text.
\par
Fig.2: Diagrams for the two-point interactions of the quark lines with the
mass shift operator $-\delta m {\tau_3}^{(1)}/2$ in place of the photon
in fig.1, but otherwise with identical topological correspondence of
figs.2(a,b,c) to figs.1(a,b,c).
\par
Fig.3: Diagrams for the e.m. self-energy of the $uud$ (proton)
configuration.
3(III) is shown in detail with full momentum markings as employed in the
text, and corresponds to quark $\# 3$ as the spectator, while the quark
lines
$\# 1$ and $\# 2$ are joined by a transverse photon line. Similarly 3(I)
and 3(II) correspond to $\# 1$ and $\# 2$ respectively as spectators in
turn.
Note that, unlike in fig.1 and fig.2, the interchange of $\# 1$ and $\# 2$
in fig.3(III) does not give a new configuration.

\end{document}